____________________________________________________________________
\\

\documentstyle[12pt]{article}
\title{{\normalsize{{\hskip 8.5cm} BIHEP-TH-94-12}}\\
{\normalsize{{\hskip 8.5cm} April,~~~~~1994}}\\
$N$-Dimensional Representations\\
of the Braid Groups $B_{N}$}

\author{Dian-Min Tong$^{1)}$, Shan-De Yang$^{1)}$ and
Zhong-Qi Ma$^{2)}$\\
\parbox[t]{15cm}{{\footnotesize {1) Department of Physics,
Jilin University, Changchun 130021, P. R. China }}\\[-2mm]
{\footnotesize {2) Institute of High Energy Physics, P. O. Box 918(4),
Beijing 100039, P. R. of China}}}}

\date{}
\begin{document}
\maketitle

\vspace{20mm}

\begin{abstract}
In this note, a new class of representations of the braid groups $B_{N}$
is constructed. It is proved that those representations contain
three kinds of irreducible representations: the trivial (identity)
one, the Burau one, and an $N$-dimensional one. The explicit form of
the $N$-dimensional irreducible representation of the braid group $B_{N}$
is given here.

\end{abstract}

\newpage
\noindent
{\bf 1. INTRODUCTION}

The braid groups $B_{N}$ consists of products of fundamental braids
$g_{i}$ and $g_{i}^{-1}$, $1\leq i \leq N-1$, satisfying:
$$g_{i}~g_{j}~=~g_{j}~g_{i},~~~~~|i-j| \geq 2 \eqno (1) $$
$$g_{i}~g_{i+1}~g_{i}~=~g_{i+1}~g_{i}~g_{i+1} \eqno (2) $$

Alexander theorem states that any oriented link can be represented
by a closed braid. Jones$^{[1]}$ discovered a generalizable
method to build a link polynomial, called Jones polynomial, from
representations of the braid groups.  Due to the similarity between
the simple Yang-Baxter equation and (2), it is easy to construct
a representation of the braid group from a solution $\breve{R}_{q}$
of the simple Yang-Baxter equation:
$$D(g_{i})~=~{\bf 1}^{(1)} \otimes \cdots \otimes {\bf 1}^{(i-1)}
\otimes \breve{R}_{q} \otimes {\bf 1}^{(i+2)} \otimes \cdots
\otimes {\bf 1}^{(N)} \eqno (3) $$

\noindent
Following Jones' method, Akutsu-Wadati$^{[2]}$ found a set of link
polynomials from representations of the braid groups constructed
by solutions of the simple Yang-Baxter equation. Jones' success
draws great attentions of mathematicians and physicists to the
braid groups.

The representations given in (3) are reducible ones.
Some works$^{[3-5]}$ have devoted to finding the irreducible
representations of the braid groups.

The braid group $B_{N}$ is an infinite discrete group with four
generators$^{[6]}$: $g_{1}$, $\Delta$ and their inverses, where
$$\Delta~=~ g_{1}~g_{2}~\cdots ~g_{N-1} \eqno (4) $$

\noindent
Under the constrain:
$$g_{i}~=~g_{i}^{-1} \eqno (5) $$

\noindent
the braid group $B_{N}$ reduces to the permutation group $S_{N}$.
Replacing the constrain (5) with (6), one obtains the Hecke algebras:
$$g_{i}~=~q^{2}~g_{i}^{-1}~+~(1-q^{2})~{\bf 1} \eqno (6) $$

\noindent
Obviously, the irreducible representations of the permutation groups and
the Hecke algebras are also the irreducible ones of the braid
groups$^{[3-5]}$. The general properties of the irreducible
representations of the braid groups were discussed in Ref. [7].
However, to find the explicit forms of the irreducible representations
completely is still an open problem.

Due to the multiplication rules (1) and (2) for the braid groups,
it is easy to obtain three irreducible representations from a known
one $D(g_{i})$ by the following ways, in addition to the usual ways,
$D^{*}(g_{i})$ and $\tilde{D}(g_{i}^{-1})$:
$$\bar{D}(g_{i})~=~\lambda~D(g_{i}) \eqno (7) $$
$$D'(g_{i})~=~D(g_{i}^{-1}) \eqno (8) $$
$$D"(g_{i})~=~\tilde{D}(g_{i}) \eqno (9) $$

\noindent
where the tilde denotes transpose, and $\lambda$ is a constant number.

Inspired by the direct product form (3),
we construct another form of representations by replacing the
direct product in (3) with the direct sum:
$$D_{m}(g_{i})~=~{\bf 1}_{i-1}~\oplus~T~\oplus ~{\bf 1}_{N-i-1} \eqno (10) $$

\noindent
where ${\bf 1}_{n}$ denotes the $n$-dimensional unit matrix, and $T$
is a $m\times m$ non-singular matrix:
$$T~=~\left(\begin{array}{cccc} T_{11}&T_{12}&\cdots &T_{1m}\\
T_{21}&T_{22}&\cdots &T_{2m}\\ \cdot &\cdot &\cdots &\cdot \\
T_{m1} &T_{m2}&\cdots &T_{mm} \end{array}\right),
{}~~~~det~T~ \neq ~0  \eqno (11) $$

\noindent
$D_{m}(g_{i})$ given in (10) is an $(N+m-2)$-dimensional representation of
the braid group $B_{N}$ if it satisfies the multiplication rules
(1) and (2). We are going to discuss the $m=2$ case in section 2,
the $m=3$ case in section 3, and the general cases in section 4.

\vspace{5mm}
\noindent
{\bf 2. $N$-DIMENSIONAL IRREDUCIBLE REPRESENTATION}

For $m=2$ case, the matrices $D_{2}(g_{i})$ given in (10) obviously satisfy
the multiplication rule (1). From the rule (2) one obtains the
conditions for the entries appearing in (11):
$$\begin{array}{l}
\left(\begin{array}{ccc}
T_{11}(T_{11}+T_{12}T_{21}) &T_{11}T_{12}(1+T_{22})
&T_{12}^{2}\\T_{11}T_{21}(1+T_{22}) &T_{12}T_{21}
+T_{11}T_{22}^{2} &T_{12}T_{22}\\T_{21}^{2} &T_{21}T_{22} &T_{22}
\end{array} \right)\\
{}~~~~~~=~\left(\begin{array}{ccc} T_{11} &T_{11}T_{12} &T_{12}^{2}\\
T_{11}T_{21} &T_{12}T_{21}+T_{11}^{2}T_{22} &
T_{12}T_{22}(1+T_{11})\\T_{21}^{2} &T_{21}T_{22}(1+T_{11}) &
T_{22}(T_{22}+T_{12}T_{21}) \end{array} \right) \end{array} $$

\noindent
namely,
$$\begin{array}{c}
T_{11}(T_{11}+T_{12}T_{21})~=~T_{11},~~~~
T_{22}(T_{22}+T_{12}T_{21})~=~T_{22} \\
T_{11}T_{22}(T_{11}-T_{22})~=~0 ,~~~~
T_{11}T_{12}T_{22}~=~0,~~~~T_{11}T_{21}T_{22}~=~0
\end{array}  \eqno (12) $$

There are four solutions to (12):

a) $T_{11}\neq 0$ and $T_{22}\neq 0$

{}From (12) we have:
$$T_{11}~=~T_{22}~=~1,~~~~~T_{12}~=~T_{21}~=~0 \eqno (13) $$

\noindent
It is the trivial (identity) representation.

b) $T_{22}=0$, $T_{11}=1-T_{12}T_{21}\neq 0$, and $T_{12}T_{21}~\neq 0$.

c) $T_{11}=0$, $T_{22}=1-T_{12}T_{21}\neq 0$, and $T_{12}T_{21} \neq 0$.

d) $T_{11}=T_{22}=0$ and $T_{12}T_{21}\neq 0$

For the solution b), $D_{2}(g_{i})$ can be simplified by a similarity
transformation X:
$$X~=~diag\left(1,~T_{21},~T_{21}^{2},~\cdots~,~T_{21}^{N-1}\right)
\eqno (14) $$
$$X^{-1}D_{2}(g_{i})X~=~{\bf 1}_{i-1}~\oplus~
\left(\begin{array}{cc} 1-t &t\\1&0 \end{array}\right)
{}~\oplus ~{\bf 1}_{N-i-1} \eqno (15) $$

\noindent
where $t=T_{12}T_{21}$. It is nothing but the Burau representation
discussed by Jones$^{[3]}$. Through another similarity transformation
$Y$, it can  be shown explicitly that the representation $D_{2}(g_{i})$
is reducible but indecomposable:
$$\begin{array}{l}
Y_{jk}~=~\left\{\begin{array}{ll} (-1)^{k-1}~~~~~&{\rm when}~~j\leq k \\
{}~~0 &{\rm when}~~j>k \end{array} \right. \\
Y_{jk}^{-1}~=~\left\{\begin{array}{ll} (-1)^{k-1}~~~~~&{\rm when}~~j
= k~~{\rm or}~~k-1 \\ ~~0 &{\rm the~else~cases} \end{array} \right.
\end{array} $$
$$\left\{ Y^{-1}X^{-1}D_{2}(g_{i})XY\right\}_{jk}~=~\left\{
\begin{array}{ll} 1~~~~~&{\rm when}~~j=k\neq i \\
-t &{\rm when}~~j=k=i \\ -t &{\rm when}~~j+1=k=i \\
-1 &{\rm when}~~j-1=k=i \\ 0 &{\rm the~else~ cases} \end{array}
\right. \eqno (16) $$

\noindent
As Jones$^{[3]}$ pointed out, removing the $N$-th row and the $N$-th
column of the representation matrices, one obtains an $(N-1)$-dimensional
irreducible representation.

For the solution c), through a similarity transformation $Z$, we obtain:
$$Z~=~diag\left(1,~T_{12}^{-1},~T_{12}^{-2},~\cdots~,~T_{12}^{1-N}\right) \eqno
(17) $$
$$Z^{-1}D_{2}(g_{i})Z~=~{\bf 1}_{i-1}~\oplus~
\left(\begin{array}{cc} 0 &1\\t&1-t \end{array}\right)
{}~\oplus ~{\bf 1}_{N-i-1} \eqno (18) $$

\noindent
where $t=T_{12}T_{21}$. In terms of (8) it also belongs to the Burau
representation.

For the solution d), through the similarity transformation $Z$ we obtain:
$$D(g_{i})~\equiv~Z^{-1}D_{2}(g_{i})Z~=~{\bf 1}_{i-1}~\oplus~
\left(\begin{array}{cc} 0 &1\\t&0 \end{array}\right)
{}~\oplus ~{\bf 1}_{N-i-1} \eqno (19) $$

\noindent
where $t=T_{12}T_{21}$. When $t=1$, the representation becomes a
special case of the Burau representation. When $t=0$, $D(g_{i})$
becomes singular.

Now, we are going to prove that this $N$-dimensional representation
with $t \neq 0$ and $1$ is irreducible. Calculate the
representation matrix of another generator $\Delta$:
$$D(\Delta)_{jk}~=~\left\{ \begin{array}{ll} 1~~~~~&{\rm when}~~
j=1~~{\rm and}~~k=N\\ t & {\rm when}~~j=k+1 \\
0 &{\rm the~else ~cases} \end{array} \right. \eqno (20) $$
$$D(\Delta^{N})~=~t^{N-1}~{\bf 1} \eqno (21) $$

\noindent
$\Delta^{N}$ is the center of the braid group $B_{N}$.

Finding the eigenvectors of $D(\Delta)$, we are able to diagonalize
it:
$$\begin{array}{l}
S_{jk}~=~t^{(j-1)/N}~\omega^{-(j-1)(k-1)},~~~~
S^{-1}_{jk}~=~N^{-1}~t^{-(k-1)/N}~\omega^{(j-1)(k-1)} \\
\omega~=~e^{i2\pi /N},~~~~~~j,~k~=~1,~2,~\cdots~,~N
\end{array} \eqno (22) $$
$$\left\{S^{-1}D(\Delta)S\right\}_{jk}~=~\delta_{jk}~t^{(N-1)/N}~
\omega^{j-1} \eqno (23) $$

\noindent
Through a direct calculation we obtain:
$$\left\{S^{-1}D(g_{1})S\right\}_{jk}~=~\delta_{jk}~-~N^{-1}~\left\{
1~-~t^{(N-1)/N}\omega^{j-1}~-~t^{1/N}\omega^{1-k}~+~\omega^{j-k}\right\}
\eqno (24) $$

\noindent
When $t=1$ the non-diagonal elements in the first row and the first column of
the matrix become vanishing so that the representation is reducible.
For the rest case ($t\neq 0$ and $1$), the $N$-dimensional representation
is obviously irreducible.

\vspace{5mm}
\noindent
{\bf 3. $m=3$ CASE}

When $m=2$ the representation of the braid group $B_{N}$ given in (10)
is the Burau representation or the $N$-dimensional representation (19).
Now we discuss the higher $m$ cases,
where the multiplication rule (1) gives some strong restrictions
on the elements of $T$. We will show that the $(N+m-2)$-dimensional
representation $D_{m}(g_{i})$ of the braid group $B_{N}$ given
in (10), $m>2$, is reducible, and it only contains three kinds of the
irreducible representations: the identity one, the Burau one and the
$N$-dimensional one.

To make the proof easier to be understood, we discuss the $m=3$
case first. From the condition (1) with $j=i+2$, we have:
$$\begin{array}{l}
T_{13}~=~T_{31}~=~0 \\
T_{12}~T_{23}~=~T_{21}~T_{32}~=~0 \\
T_{23}~(T_{11}~-~1)~=~T_{32}~(T_{11}~-~1)~=~0 \\
T_{12}~(T_{33}~-~1)~=~T_{21}~(T_{33}~-~1)~=~0 \end{array} \eqno (25) $$

{}From (2) and (25) we have:
$$\begin{array}{ll}
D_{3}(g_{i})~=~D_{2}(g_{i})~\oplus~1,~~~~&{\rm when}~~T_{23}~=~T_{32}~=~0\\
D_{3}(g_{i})~=~1~\oplus~D_{2}(g_{i}), &{\rm when}~~T_{12}~=~T_{21}~=~0
\end{array} \eqno (26) $$

When $T_{12}\neq 0$ and $T_{32}\neq 0$, from the
condition (2) and through an appropriate similarity transformation,
we obtain:
$$D_{3}(g_{i})~=~{\bf 1}_{i-1}~\oplus~\left(\begin{array}{ccc}
1&-t&0\\0&-t&0\\0&-1&1 \end{array}\right)~\oplus~{\bf 1}_{N-i-1}
 \eqno (27) $$

\noindent
It is the typical Burau representation$^{[3]}$ that is reducible and
contains two identity representations.

When $T_{21}\neq 0 $ and $T_{23}\neq 0$, the representation is
the same as (27) except for a transpose. Due to (9) the conclusion
is the same.

\vspace{5mm}
\noindent
{\bf 4. $m\geq 3$ CASE}

Substituting (10) into (1) we obtain the following useful conditions
for the $m\times m$ matrix $T$:
$$\begin{array}{ll}
T_{ab}~=~0, &{\rm if}~~|a-b|\geq 2 \\
T_{a(a+1)}~T_{b(b+1)}~=~T_{a(a-1)}~T_{b(b-1)}~=~0,~~~~&{\rm if}~~
a\neq b \\
T_{a(a+1)}~(T_{11}~-~1)~=~T_{(a+1)a}~(T_{11}~-~1)~=~0
,~~~~&{\rm if}~~a\neq 1 \\
T_{a(a-1)}~(T_{mm}~-~1)~=~T_{(a-1)a}~(T_{mm}~-~1)~=~0
,~~~~&{\rm if}~~a\neq m \\
T_{12}~T_{m(m-1)}~=~T_{21}~T_{(m-1)m}, &{\rm when}~~ m\geq 4
 \end{array} \eqno (28) $$

Thus, if $T_{12}=T_{21}=0$, we have:
$$D_{m}(g_{i})~=~1~\oplus~D_{m-1}(g_{i}) \eqno (29) $$

\noindent
If $T_{12}\neq 0$ or $T_{21}\neq 0$, we have $T_{m(m-1)}=T_{(m-1)m}=0$,
and:
$$D_{m}(g_{i})~=~D_{m-1}(g_{i})~\oplus~1 \eqno (30) $$

Therefore, the $(N+m-2)$-dimensional representation $D_{m}(g_{i})$
of the braid group $B_{N}$ given in (10), $m\geq 3$, is reducible,
and it only contains three kinds of the irreducible representations:
the identity one, the Burau one and the $N$-dimensional one given in (19).

\vspace{5mm}
{\bf Acknowledgments}. This work was supported by the National
Natural Science Foundation of China and Grant No. LWTZ-1298 of
Chinese Academy of Sciences.


\begin{thebibliography}{99}
\bibitem{1} V. F. R. Jones, {\it Bull. Am. Math. Soc}. {\bf 12}(1985)103.

\bibitem{2} Y. Akutsu and M. Wadati, {\it J. Phys. Soc. Jpn}.
{\bf 56}(1987)3039.

\bibitem{3} V. F. R. Jones, {\it Ann. Math}. {\bf 126}(1987)335.

\bibitem{4} D. Kazdan and G. Lusztig, {\it Invent. Math}. {\bf 53}(1979)165;
A. Gyoja, {\it Osaka J. Math}. {\bf 23}(1986)841; H. Wenzl, {\it
invent. Math}. {\bf 92}(1988)349.

\bibitem{5} Dian-Min Tong, Cheng-Jiu Zhu and Zhong-Qi Ma, {\it J.
Math. Phys}. {\bf 33}(1992)2260.

\bibitem{6} Zhong-Qi Ma, {\it Yang-Baxter Equation and Quantum Enveloping
algebras}, World Scientific, Singapore, 1993.

\bibitem{7} Zhong-Qi Ma, Dian-Min Tong and Bin Zhou, {\it Commun.
Theor. Phys}. {\bf 18}(1992)369.

\end{thebibliography}
\end{document}